\documentclass[aps,prl,twocolumn,superscriptaddress]{revtex4-1}

\usepackage{graphicx}

\begin{document}
\title{Direct observation of Oersted-field-induced magnetization dynamics in magnetic nanostripes}
\author{V. Uhl\'\i\v r}
\affiliation{Institut N\'{e}el, CNRS and UJF, BP166, 38042 Grenoble,
France} \affiliation{Institute of Physical Engineering, Brno
University of Technology, 61669 Brno, Czech Republic}
\author{S.~Pizzini}
\affiliation{Institut N\'{e}el, CNRS and UJF, BP166, 38042 Grenoble,
France}
\author{N.~Rougemaille}
\affiliation{Institut N\'{e}el, CNRS and UJF, BP166, 38042 Grenoble,
France}
\author{V.~Cros}
\affiliation{Unit\'{e} Mixte de Physique CNRS/Thales, Route
d\'{e}partementale 128, 91767 Palaiseau cedex,
France}
\author{E.~Jim\'{e}nez}
\affiliation{Dpto. F\'{i}sica de la Materia Condensada, Instituto
``Nicol\'{a}s Cabrera" and IMDEA-Nanociencia, Campus Universidad
Aut\'{o}noma de Madrid, 28049 Madrid, Spain}
\author{L.~Ranno}
\affiliation{Institut N\'{e}el, CNRS and UJF, BP166, 38042 Grenoble,
France}
\author{O.~Fruchart}
\affiliation{Institut N\'{e}el, CNRS and UJF, BP166, 38042 Grenoble,
France}
\author{M. Urb\'{a}nek}
\affiliation{Institute of Physical Engineering, Brno University of
Technology, 61669 Brno, Czech Republic}
\author{G.~Gaudin}
\affiliation{SPINTEC, UMR8191, CEA/CNRS/UJF/GINP, INAC, 38045
Grenoble, France}
\author{J.~Camarero}
\affiliation{Dpto. F\'{i}sica de la Materia Condensada, Instituto
``Nicol\'{a}s Cabrera" and IMDEA-Nanociencia, Campus Universidad
Aut\'{o}noma de Madrid, 28049 Madrid, Spain}
\author{C.~Tieg}
\affiliation{ESRF, BP200, 38043 Grenoble, France}
\author{F.~Sirotti}
\affiliation{Synchrotron SOLEIL, L'Orme des Merisiers, Saint-Aubin,
91192 Gif-sur-Yvette, France}
\author{E.~Wagner}
\affiliation{Institut N\'{e}el, CNRS and UJF, BP166, 38042 Grenoble,
France}
\author{J.~Vogel} \email{jan.vogel@grenoble.cnrs.fr}
\affiliation{Institut N\'{e}el, CNRS and UJF, BP166, 38042 Grenoble,
France} 

\begin{abstract}
We have used time-resolved x-ray photoemission electron microscopy
to investigate the magnetization dynamics induced by nanosecond
current pulses in NiFe/Cu/Co nanostripes.
A large tilt of the NiFe magnetization in the direction transverse to the stripe is observed during the pulses. We show that this effect cannot be quantitatively understood from the amplitude of the Oersted field and the shape anisotropy.
High frequency oscillations observed at the onset of the pulses are attributed to
precessional motion of the NiFe magnetization about the effective field. We discuss the possible origins of the large magnetization tilt and the potential implications of the static and dynamic effects of the Oersted field on current-induced domain wall motion in such stripes.
\end{abstract}

\maketitle

The possibility to manipulate the magnetic configuration of
nanostructures by using electrical currents is a recent, exciting
development in spintronics. Electrical currents can affect the
magnetization of magnetic nanostructures both through the charge and the
spin of the conduction electrons. In recent years it has been shown that Spin-Transfer Torque
(STT) \cite{Berger1984,Slonczewski1996} and Rashba spin-orbit torque effects
\cite{Miron2010} act on the magnetization, in addition to the classical Oersted magnetic field
(H$_{Oe}$). In general, the combination of these effects should be taken into account in the description of the magnetization dynamics during the application of a current pulse.  For
instance, it was shown that the contribution of the Oersted field and not only STT
is needed to explain the magnetization reversal in trilayered
pillars induced by a current flowing perpendicular to the plane of
the layers \cite{Ito2007,Acremann2006}. For in-plane currents,
H$_{Oe}$ has been invoked to explain magnetization reversal in mesoscopic
NiFe/Cu/Co/Au bars \cite{Morecroft2007} and the resonant depinning of
constricted domain walls (DWs) in NiFe/Cu/Co trilayers \cite{Metaxas2010}.

Several studies of the effect of current pulses on the magnetization of nanostripes, mainly concerning current-induced domain wall motion (CIDM), have been based on the observation of the domain structure
 before and after the application of a current pulse
\cite{Yamaguchi2004,Klaui2005}. However, the effect of the Oersted
field on the magnetization can only be investigated by direct, dynamic
observations \textit{during} the current pulses. This has been achieved in this work, using
time-resolved x-ray magnetic circular dichroism combined with
photoemission electron microscopy
(XMCD-PEEM). Our results show that the current-induced field during nanosecond pulses causes both quasi-static and precessional effects on
the NiFe magnetization. These effects may
contribute to the increased efficiency of current-induced domain
wall motion observed in such trilayers \cite{Grollier2003,Pizzini2009,Uhlir2010}.

Stacks of Cu(2nm)/Ni$_{80}$Fe$_{20}$(5nm)/Cu(5nm)/Co(5nm)/ CoO(6nm) deposited on highly resistive Si(100) ($\rho > 300~\Omega$.cm) were patterned in 400 nm wide zigzag stripes, with angles of 90$^{\circ}$ and 13 $\mu$m long
straight sections, combining electron beam lithography and ion-beam
etching. Contact
electrodes made of Ti/Au were subsequently deposited using
evaporation and a lift-off technique. Prior to the XMCD-PEEM measurements, most of the 2~nm Cu protective layer was removed in-situ using Ar-bombardment, to increase the XMCD signal of the NiFe layer.

XMCD-PEEM measurements were
performed at the synchrotron SOLEIL (TEMPO beamline),
using a Focus IS-PEEM. The magnetic configuration in the NiFe
layer was imaged by measuring the Ni XMCD intensity, tuning the x-ray energy to the Ni $L_3$
absorption edge (852.8 eV). To optimize the magnetic contrast, the difference between
two consecutive images obtained with $100\%$ left- and
right-circularly polarized x-rays was computed. For each circular polarization, 60 images of 0.5 s were summed, after correction for possible image drifts.

Temporal resolution was obtained by synchronizing nanosecond current pulses applied to the nanostripes with the SOLEIL 8-bunch mode, where photon bunches arrive at the sample with a repetition rate of 6.77 MHz. The temporal evolution of the magnetic
configuration in the nanostripes was obtained by recording images for different delays between
the current and photon pulses
\cite{Sirotti2000,Bonfim2001,Vogel2003,Schonhense2006}. If events are reproducible and reversible for each current pulse, the temporal resolution of this pump-probe technique is limited only by the duration of the photon pulses (50-60 ps). The total acquisition time of 1 minute for each XMCD-image implies that sequences of about $4 \times 10^{8}$
current (pump) and photon (probe) pulses were averaged. In order to avoid electrical discharges, the
voltage between the sample and the objective lens of the PEEM was set
to 5.4 keV instead of the nominal 12 keV, limiting the spatial
resolution to about 0.6 $\mu$m.

\begin{figure}[ht!]
\includegraphics{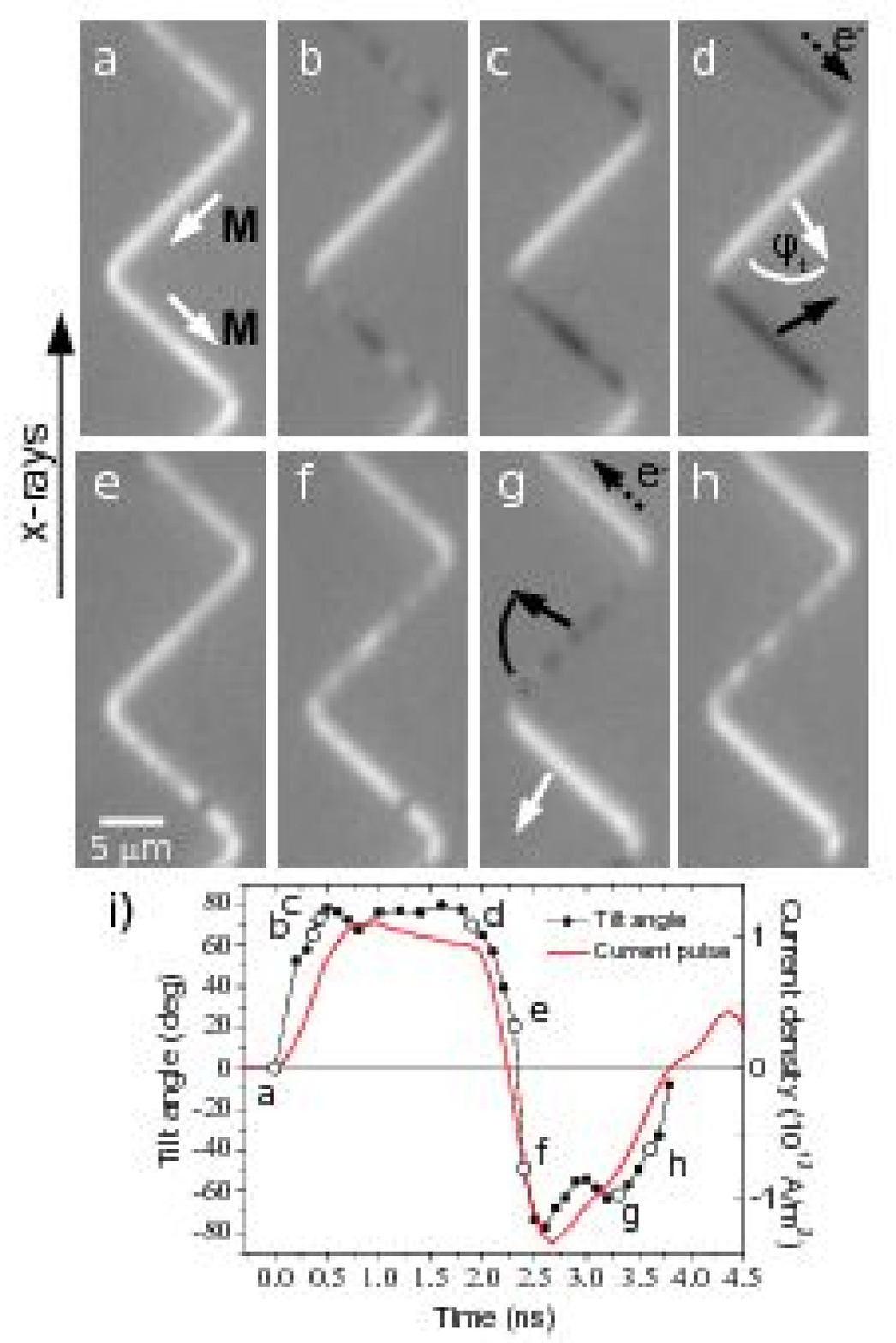}
\caption{\label{fig:movie} (color online) Time-resolved XMCD-PEEM
images of the NiFe layer of a 400 nm wide nanostripe at time delays of
(a) 0 ns, (b) 0.35 ns, (c) 0.45 ns, (d) 1.9 ns, (e) 2.3
ns, (f) 2.4 ns, (g) 3.3 and (h) 3.6 ns with respect to the beginning
of the positive part of the bipolar current pulse \cite{Website}. These delays are
indicated on the bipolar pulse plotted in (i), together with the
magnetization tilt angle $\varphi_t$. The oscillations in
$\varphi_t$ at the beginning of the positive and negative parts of
the pulse indicate magnetization precession about H$_{Oe}$.}
\end{figure}

Figure 1 shows a series of XMCD-PEEM images of the NiFe magnetization acquired during the application of bipolar current pulses
[Fig.~\ref{fig:movie}]. The positive/negative part of the pulse is
about 2 ns/1 ns long, with a maximum amplitude of $+7$ mA/$-9$ mA.
The latter value corresponds to a current density of $1.5 \times
10^{12}$ A/m$^2$ assuming a homogeneous current distribution in the
stack.
Before and after the current
pulses, the magnetization is aligned along the stripe axis and no domain
walls are present, leading to an almost homogeneous XMCD intensity
[Fig.~\ref{fig:movie}(a)]. During the current pulses, the NiFe
magnetization tilts away from the wire axis, with a tilt angle
$\varphi_t$. This tilt is anti-clockwise for a positive and
clockwise for a negative current direction, as can be inferred from
the magnetic contrast in the differently oriented sections of the
nanowire [Figs.~\ref{fig:movie}(d) and (g)]. The approximate magnetization directions in two of the wire sections are indicated before the current pulses (a), and at the end of the plateau of the positive (d) and negative part of the pulses (g). In (d) and (g) also the electron flow directions are indicated. The delays between the beginning of the current pulse and the photon pulses at which the images were acquired are shown in Fig.~\ref{fig:movie}(i) \cite{Website}.

In order to obtain the tilt angle $\varphi_t$ as a function of time during the current pulses [Fig.~\ref{fig:movie}(i)], the normalized XMCD-intensity in the bends of the stripe was determined from the XMCD-PEEM images. The XMCD intensity is proportional to the cosine of the angle between the incoming x-rays and the local magnetization, thus for the bends $I_{\text{XMCD}} \propto M\cos(\varphi_t)\cos(\alpha)$, where M is the magnetization and $\alpha$ is the angle between the x-ray incidence direction and the sample surface. This angle is constant (25$^\circ$) and we also suppose the magnetization vector has a constant amplitude. No change of the magnetic contrast due to current-induced heating was observed. At zero current, the magnetization is parallel to the stripe axis and thus $\varphi_t$ = 0$^\circ$, giving $I_{\text{XMCD}} = Mcos(\alpha) = I_0$. Then $\varphi_t$ can be determined from the different images by $\varphi_t = arccos(I_{\text{XMCD}}/I_0)$.

The NiFe magnetization tilt induced by the transverse Oersted field is surprisingly large, with a value of about $75^\circ$ at the end of the positive part of the pulse. For a soft magnetic
material such as NiFe, the magnetization direction in a nanostripe is
mainly determined by magnetostatic effects, which favor
magnetization along the stripe axis. For a 5 nm thick, 400 nm wide
stripe the transverse demagnetizing factor is about 0.023
\cite{Aharoni1998}. In a first approximation, this would mean that a transverse field of $0.023 \times \mu_0 M_S \times \sin
75^{\circ} = 22$ mT (with $\mu_0 M_S$ = 1 T for
permalloy) would be required  to obtain $\varphi_t$ = 75$^\circ$.

The Oersted field inside a stripe with rectangular cross-section is
given by $B_x=\mu_0Jz$, where J is the current density and z is the
distance from the stripe axis. A current of $+7$ mA corresponds to a current density of $1.17 \times
10^{12}$ A/m$^2$, yielding an average field acting on the NiFe magnetization of H$_{Oe}$= 7.4 mT if we assume a homogeneous
current distribution over the NiFe/Cu/Co trilayer structure, and 11 mT for a current flowing entirely through the Cu and
Co layers. With a field of 11 mT, the expression given above yields a
$\varphi_t$ of only 28$^\circ$ instead of the
observed 75$^\circ$.

\begin{figure}[ht!]
\includegraphics*[bb= 214 420 406 545]{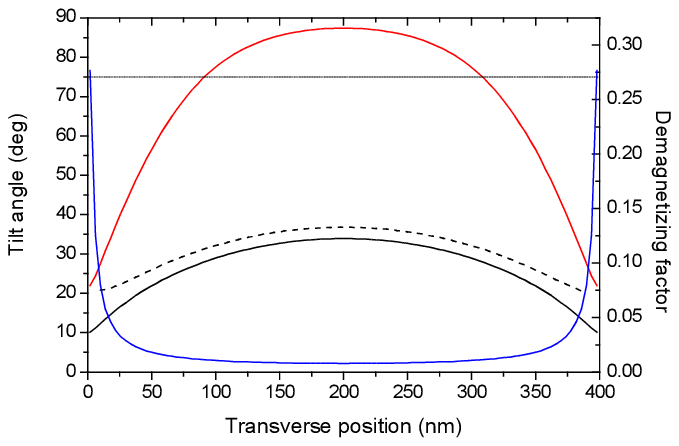}
\caption{\label{fig:simul} (color online) NiFe magnetization tilt angle as a function of the transverse position inside a 400 nm wide stripe, obtained using the OOMMF code, for a transverse Oersted magnetic field of 7.4 mT. The experimental NiFe tilt angle is indicated with a dashed line. Different cases were considered : a single, 5 nm thick NiFe layer without edge roughness (black), a 5 nm think NiFe with a random edge roughness of 4-8 nm (dotted black), and a NiFe(5nm)/Cu(5nm)/Co(5nm) trilayer (red). The demagnetization factor along the wire is also shown (blue, right y-axis)}
\end{figure}

The most likely origin of the discrepancy between the observed and expected tilt
angles is an overestimation of demagnetizing effects. The value of $\varphi_t = 28^\circ$ is obtained assuming that the tilt is homogeneous over the stripe width.
 In reality, the demagnetizing effect is much smaller in the center than at the edges of the stripe, leading to a larger tilt angle in the center. We carried out micromagnetic simulations using the OOMMF code \cite{OOMMF} to obtain the magnetization profile
of a 400 nm wide, 5 nm thick layer of Ni$_{80}$Fe$_{20}$ and for NiFe(5nm)/Cu(5nm)/Co(5nm) trilayers under an Oersted field of 7.4 mT. The results obtained for NiFe (Co) using an exchange constant A of $1 \times 10^{-11}$ J/m ($3 \times 10^{-11}$ J/m), a spontaneous magnetization $\mu_0M_S$ of 1T (1.76T) and a vanishing  magnetic anisotropy constant K, are shown in Fig.~\ref{fig:simul}. The blue (black) continuous line shows the demagnetizing factor ($\varphi_t$) for a single Py layer, as a function of transverse position. The average tilt angle is 26.6$^\circ$ with a maximum of 32$^\circ$ in the center of the stripe. As shown by previous studies, edge roughness can lead to a decrease of the transverse demagnetizing factor of several tens of percent \cite{Cowburn2000}. The simulated magnetization profile obtained by adding a random lateral roughness of 4-8 nm (1-2 grid cells) at the stripe edges (dotted black line) leads only to a slight increase of the average tilt angle (to about 30$^\circ$). Edge roughness is therefore not sufficient to explain the large experimental tilt. A better quantitative agreement with experiments can be obtained by taking into account the presence of the Co layer. Magnetostatic
interactions between the NiFe and Co layers can significantly
decrease the transverse demagnetizing effects with respect to single
NiFe wires. Part of the magnetic charges on the edges of the NiFe
layer can be compensated by mirroring effects on the edges of the Co
layer, as shown by micromagnetic simulations
\cite{Ndjaka2009}. Moreover, if the current is centered in the Cu
layer the Co magnetization tilt induced by H$_{Oe}$ will be opposite
to the one induced in the NiFe layer, further increasing the
compensating effect of the Co magnetic charges. The average tilt angle obtained for the NiFe layer in the case of a NiFe/Cu/Co trilayer
is around 69$^\circ$, close to the experimental value, with a Co tilt angle (not shown) of about -42$^\circ$. The NiFe magnetization tilt strongly depends on the Co tilt angle. In the simulations of Fig.~\ref{fig:simul}, the magnetic anisotropy in the Co layer was taken to be zero, which is justified by the polycrystalline nature of the Co leading to the absence of an in-plane uniaxial anisotropy before patterning. A uniaxial Co anisotropy along the stripe axis of 50 kJ/m$^3$ would lead to a Co tilt angle of only -9$^\circ$, and a NiFe tilt of 42$^\circ$.

\begin{figure}[ht!]
\includegraphics*[bb= 247 439 362 534]{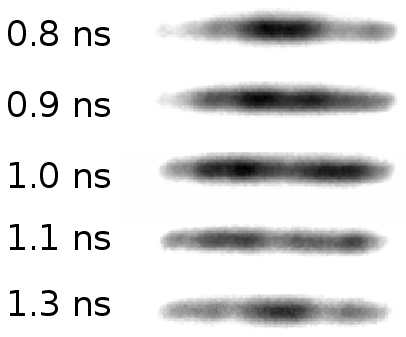}
\caption{\label{fig:spinwaves} Time-resolved XMCD-PEEM images of the lower, 13
$\mu$m long section of the nanostripe, taken at the
indicated delays after the beginning of the positive part of the
current pulse. Spatio-temporal variations of the XMCD contrast at a frequency of about 2
GHz are visible~\cite{Website}.}
\end{figure}

At the onset of the current pulse, fast oscillations of the magnetization are observed in the time-resolved images [Fig.~\ref{fig:movie}(i)]. These oscillations are due to the precession of the magnetization about the effective field. Figure~\ref{fig:spinwaves} shows contrast-enhanced XMCD-PEEM images of the bottom section of the nanostripe of Fig.~\ref{fig:movie}. Inhomogeneities in the dipolar interactions with the Co layer and in edge roughness lead to inhomogeneous magnetic contrast in the sections of the spin-valve nanowire. Different parts of the nanowire oscillate with different initial phases of the precessional motion. The exchange interaction between the different parts, however, leads to spatio-temporal variations of the magnetic contrast that resemble spin waves. The oscillatory and propagative nature of these contrast variations is more clearly visible in the accompanying movie \cite{Website}. The excitation of spin-waves by the
Oersted field in spin-valve trilayers was predicted by Kim et al.
\cite{Kim2007} and spin-wave-like features were observed using
Lorentz microscopy on 30 nm thick NiFe nanostripes upon current
injection \cite{Togawa2008}. Further micromagnetic simulations are necessary to understand these oscillations quantitatively, but our results show that time-resolved XMCD-PEEM is a very suitable technique to observe such magnetization
oscillations.

In quasi-static measurements performed on similar nanostripes we have
observed that current pulses with a density above $1.5-2 \times
10^{12}$ A/m$^2$ can induce nucleation of reversed domains in
initially saturated nanostripe sections \cite{Uhlir2010}. Our present results suggest that the precession of the
magnetization about H$_{Oe}$ is possibly at the origin of this local magnetization reversal, similar to the magnetization reversal \cite{Schumacher2003a}
induced by transverse magnetic field pulses in magnetic nanostructures.

The magnetization tilt induced by the Oersted field and amplified by magnetostatic interactions should also have an influence on current-induced domain wall motion in such trilayers
\cite{Grollier2003,Pizzini2009,Uhlir2010}. The amplified Oersted field
might stabilize transverse domain walls having their
magnetization parallel to H$_{Oe}$, like it was observed for field-induced domain wall motion in trilayer
nanostripes in the presence of a transverse magnetic field \cite{Bryan2008,Glathe2008b}.

In conclusion, we provide direct, time-resolved microscopic
evidence of the effect of current-induced fields on the magnetic
configuration of magnetic nanostripes. We
show that the combination of Oersted fields and strong dipolar interactions that may exist in nanostripes comprised of several metallic layers produce and amplify a large tilt of the magnetization. The quasi-static and precessional effects induced by the amplified Oersted
field should be carefully considered when current pulses are applied to magnetic wires with different metallic layers, for instance to study current-induced domain wall motion. On the other hand, the effect
of the Oersted field on magnetization reversal or magnetic domain
wall motion in future spintronic devices can be tailored by tuning the thickness of the
different metallic layers.

We acknowledge the invaluable technical and experimental help of
P.~Perrier, D.~Lepoittevin, L.~Delrey, S.~Pairis,
T.~Fournier, A.~Hrabec, M.~Bonfim and W.~Wernsdorfer. We thank A.~Anane,
J.~Grollier and R.~Mattana for experimental help and useful
discussions. We thank the European Synchrotron Radiation Facility
(ESRF), and in particular the staff of beamline ID08, where several
preliminary experiments were carried out. Nanofabrication was
performed at the `Plateforme de Technologies Avanc\'{e}es' and at
the Institut N\'{e}el/CNRS `Nanofab' facility, both in Grenoble.
E.J. and J.C. acknowledge financial support through projects
HF2007-0071, S2009/MAT-1726, and CSD 2007-00010. V.U. was
financially supported by grants No. MSM0021630508, No. KAN400100701
and No. 2E13800101-MSMT, and by the project INGO No. LA287 of the
Czech Ministry of Education. This work was partially supported by
the ANR-07-NANO-034 `Dynawall'.

\bibliographystyle{apsrev}

\end{document}